\documentclass[prd,12pt]{revtex4}
\usepackage{amsmath,amssymb} 
\usepackage[dvips]{graphicx} 
\usepackage{graphics}
\usepackage{epsfig}

\newcommand\nn{\nonumber}
\newcommand\ba{\begin{eqnarray}}
\newcommand\ea{\end{eqnarray}}
\newcommand\bg{\begin{gather}}
\newcommand\eg{\end{gather}}
\newcommand{\br}[1]{\left( #1 \right)}
\newcommand{\brs}[1]{\left[ #1 \right]}
\newcommand{\brf}[1]{\left\{ #1 \right\}}
\newcommand{\brm}[1]{\left| #1 \right|}

\newcommand{\cm}{~\mbox{cm}}

\begin{document}

\title{Two-pion production in electron-polarized proton scattering}

\author{A.I. Ahmadov}
\email{ahmadov@theor.jinr.ru}
\affiliation{JINR-BLTP, 141980 Dubna, Moscow region, Russian Federation}
\affiliation{Institute of Physics, Azerbaijan National Academy of Sciences,
Baku, Azerbaijan}

\author{E.A. Kuraev}
\email{kuraev@theor.jinr.ru}
\affiliation{JINR-BLTP, 141980 Dubna, Moscow region, Russian Federation}

\author{Yu.M. Bystritskiy}
\email{bystr@theor.jinr.ru}
\affiliation{JINR-BLTP, 141980 Dubna, Moscow region, Russian Federation}

\begin{abstract}
    The process of two pion production in the electron-polarized proton scattering is
    investigated. In the Weizs\"acker-Williams approximation the differential spectral
    distributions and the spin-momentum correlations are considered.
    The spin correlation effects caused by $\rho$-meson widths
    are estimated to be of an order of several per cent.
    Both channels of the $\pi^+\pi^-$ and $\pi^+\pi^0$ creation are considered.
    The effects of intermediate excited baryons are not considered.
    The spectral distributions on pion energy fractions in polarized
    and unpolarized cases are presented analytically and numerically.
\end{abstract}

\maketitle

\section{Introduction}
\label{Introduction}

In the Semi-Inclusive Deep Inelastic Scattering (SIDIS) setup (see papers
\cite{jaffe, radici, bijnens, kopytin,korotkov, radici1, bacchetta, brull, jaffe1, baranov, aa1})
with the production of ($2\pi$) states in the $ep$ - scattering we studied azimuthal correlation
between the direction of polarization of the proton and the transverse momentum
of the pion. In the SIDIS description ($ep \to eN\pi\pi$) in the kinematics of the forward electron
scattering there are two different mechanism of production $2\pi$.
One of them (so called "two photon mechanism") contains the sub-process $\gamma\rho \to 2 \pi$
with the emission of almost real photon by the electron and the $\rho$ -meson emission by the nucleon.
The other is connected with the "Compton mechanism". It contains the sub-process when the photon directly
interacts with the proton and creates a $\rho$ meson with its subsequent decay into a pair of pions.

The kinematics considered corresponds to the fragmentation region of the proton. Besides we restrict ourselves
to the case when the invariant mass of the the final state $N 2\pi$ is small enough to exclude the baryon
resonances in the intermediate state.

We consider two channels with the creation of $\pi^+\pi^-$ and $\pi^+\pi_0$ states.

In this kinematical region we use the approach of low-energy theory of nucleon-pion interaction
\cite{Gaziorovich,Gourdin,Meissner,Witten,Kaymakcalan}.

The alternative way is associated with applying the perturbation theory of QCD. It must take into
account the contributions arising from the higher twist operators.
Another is to use the CHPT theory. Here, both these theories are not dealt with.

In this paper, we discuss the possibility of separate study of these two
different mechanisms of SIDIS, based on the analysis of Dalitz
distributions in the system $(p \pi\pi), (n \pi\pi)$.

To do this, we consider the ratio of polarized and unpolarized contributions.

The Dalitz-plot distributions in the case of unpolarized initial proton
are considered as well.

We suggest the independent method of studying SIDIS by measuring
the Dalitz distribution of the final system $p\pi^+\pi^-,\,\,n\pi^+\pi^0$.

Different mechanisms of the creation of $\pi^+\pi^- (\pi^+\pi^0)$ final state are
considered in the case of scattering of the electron on the polarized proton.

The relevant Feynman diagrams are presented in Fig1.
\begin{figure}
\includegraphics[width=0.8\textwidth]{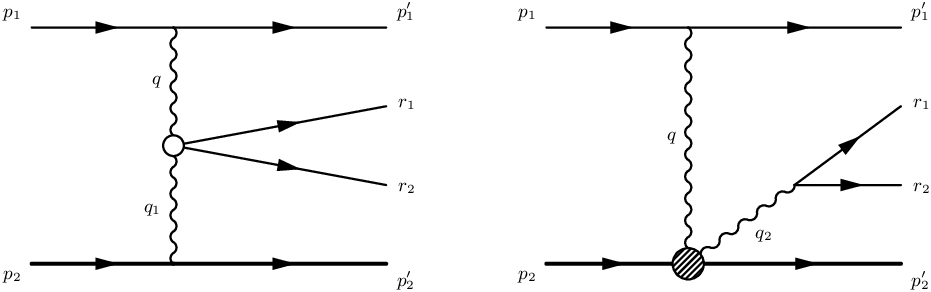}
\caption{Two mechanisms production of $2\pi$ in electron proton scattering.}
\label{Fig1}
\end{figure}

The near-forward electron scattering kinematics (first diagram in Fig.~\ref{Fig1})
provides the Weizs\"acker-Williams (WW) enhancement of the cross section \cite{Akhiezer:1965}
by a factor $L=\ln\br{s^2/\br{M_p^2 m_e^2}}\sim 20$, common for both mechanisms.
Besides there is the Compton mechanism (see second diagram in Fig.~\ref{Fig1}) which contains
Breit-Wigner enhancement factor associated with the intermediate $\rho$-meson state decaying into a pion.
It permits one in principle to distinguish the contributions of two mechanisms.

The specific feature of the Compton mechanism - is the possibility to study the one-spin correlations
of the proton spin direction with the momenta of the pion in the final state.
Due to a rather large width of the $\rho$-meson this correlation can presumabely be measured
\cite{Sivers1,Sivers2}.

Both experimental set-ups the case of the fixed proton target (HERMES facility) and
the case of colliding beams (ZEUS facility).
These reasons are the motivation of this paper.

Our paper is organized as follows. In the first chapter, we describe the kinematics of forward scattering
to be applied to the process studied and infer the form of the differential cross section.
In the second part, the channel  with production of a pair of charged pions is considered.
In the next parts, the channel with $\pi^+\pi^0$ production is considered and the results of
numerical analysis of unpolarized and polarized parts of the cross sections are presented.

To study the processes
\ba
e^-(p_1)+p(p) &\to&
e^-(p_1')+N(p')+\pi_i(r_1)+\pi_j(r_2); \,\,\,\pi_i=\pi_+,\pi_-,\pi_0,
\ea
in the kinematics of the near forward electron scattering
\ba
s=2pp_1>>|q^2|, \,\,q=p_1-p_1', \,\,p^2=p^{'2}=M^2, \,\,r_1^2=r_2^2=m^2, \nn \\
s_1=(p'+r_1+r_2)^2-M^2<<s,
\ea
it is convenient to use the basis of two light-cone vectors built from the momenta of the
initial particles and use the Sudakov parametrization of the momenta
of other particles \cite{ECHAJA}
\ba
&&\tilde{p}_1=p_1-p\frac{m_e^2}{s}; \qquad \tilde{p}=p-p_1\frac{M^2}{s}, \nn \\
&&\tilde{p}_1^2=O\br{m_e^4\frac{M^2}{s^2}}; \qquad \tilde{p}^2=O\br{m_e^4\frac{M^4}{s^2}},\nn \\
&&2\tilde{p}\tilde{p}_1=s; \qquad 2p\tilde{p}=M^2; \qquad 2p_1\tilde{p}_1=m_e^2.
\ea
Without loss of the accuracy we use below $\tilde{p}^2=\tilde{p}_1^2=0$.
Accept now the Sudakov parametrization
\ba
q&=&\alpha \tilde{p}+\beta\tilde{p}_1+q_\bot; \nn \\
r_i&=&x_i\tilde{p}+\beta_i\tilde{p}_1+r_{i\bot}; \nn \\
p_1'=p_1-q; p'&=&x\tilde{p}+\beta\tilde{p}_1+q_\bot,
\ea
with $q_{i\bot} p=q_{i\bot} p_1=0$ and $q_{i\bot}^2=-\vec{q}_i^2$. The conservation law reads as
\ba
x+x_1+x_2=1; \vec{q}=\vec{p}'+\vec{r}_1+\vec{r}_2;\beta=\beta'+\beta_1+\beta_2-\frac{M^2}{s}.
\ea
Using the on mass shell condition of the scattered electron $p_1^{'2}-m_e^2=0$ one finds
\ba
q^2=-\frac{1}{1-\beta}[\vec{q}^2+q_0^2], q_0^2=m_e^2\beta^2=m_e^2(\frac{s_0}{s})^2, \nn \\
s_0=s\beta=\frac{1}{x_1}[\vec{r}_1^2+m^2]+\frac{1}{x_2}[\vec{r}_2^2+m^2]+\frac{1}{x}[(\vec{r}_1+\vec{r}_2)^2+(1-x)M^2].
\ea
Here we also use the on mass shell conditions for pions.

The matrix element in the Born approximation has the form
\ba
M^{ep \to eN2\pi}=\frac{4\pi\alpha G}{q^2}
\brs{\bar{u}(p_1')\gamma_\mu u(p_1)} J_\nu(q) g^{\mu\nu},
\ea
where $J_\nu(q)$ is the hadron current
\ba
J_\nu(q)=\bar{u}(p')O_\nu u(p),
\ea
which obeys the current conservation condition
$J_\lambda(q)q^\lambda=0$,
operator $O_\nu$ will be defined below.
The quantity $G$ is a product of $\rho$-meson coupling
constants $G=g_{\rho\pi\pi}g_{\rho NN}\approx 20$.

Significant simplification comes from the Gribov form of the metric
tensor $g^{\mu\nu}$ in the photon propagator \cite{ECHAJA}:
\ba
g^{\mu\nu}=g_\bot^{\mu\nu}+\frac{2}{s}\brs{\tilde{p}^\mu\tilde{p}_1^\nu+
\tilde{p}^\nu\tilde{p}_1^\mu} \approx
\frac{2}{s}\tilde{p}^{\mu}\tilde{p_1}^{\nu}.
\ea
In the kinematics we are considering, the only
$(2/s)\tilde{p}^\mu\tilde{p}_1^\nu$ component of this tensor is relevant.
Other terms give the contributions which are suppressed by the factors
\ba
O\br{\frac{M^2}{s}} \ll 1.
\ea
We also systematically omit the terms of an order of
\ba
O\br{\frac{m^2}{M^2}}\sim0.02.
\ea
The terms of this order of magnitude (which are systematically omitted below)
determine the accuracy of our consideration.

Thus, the matrix element has the form:
\ba
M^{ep\to eN2\pi}&=&\frac{8\pi s\alpha}{q^2}N_e M_h^i,
\qquad
i=\pi^+\pi^-, \pi^+ \pi^0,\\
M_h^i&=&\frac{1}{s}\bar{u}(p')p_1^\nu O_\nu(q)u(p,a), \nn \\
N_e&=&\frac{1}{s}\brs{\bar{u}(p_1')\gamma_\mu u(p_1)}\tilde{p}^\mu. \nn
\ea
It is easy to see that
\ba
\sum|N_e|^2=2,
\ea
and that the quantity $N_h$ does not depend on $s$ in the large $s$ limit.

Let us now parameterize the phase volume of the final particles:
\ba
d\Gamma=(2\pi)^{-8}\frac{d^3p_1'}{2E_1'}\frac{d^3p'}{2E'}
\frac{d^3q_1}{2E_1}\frac{d^3q_2}{2E_2}\delta^4(p+p_1-p'-p_1'-q_1-q_2).
\ea
We use below the on-mass shell relation for pions
\ba
\frac{d^3r_i}{2r_{0i}}=d^4r_i\delta\br{r_i^2-m^2}=
\frac{s}{2}dx_i d\alpha_i d^2\vec{r}_i
\delta\br{sx_i\beta_i-\vec{r}_i^2-m^2}=\frac{1}{2x_i}dx_i d^2\vec{r}_i, i=1,2.
\ea
Introducing the additional unit factor
\ba
1 = d^4q~\delta^4\br{p_1-p_1'-q}=\frac{s}{2}d\alpha~ d\beta~
d^2\vec{q}~ \delta^4\br{p_1-p_1'-q}
\ea
we put the phase volume to the form
\ba
d\Gamma=\br{2\pi}^{-8}\frac{dx_1 dx_2}{8s x x_1 x_2}
d^2\vec{q} d^2\vec{r}_1^2 d^2\vec{r}_2.
\ea

One can see that the module of the matrix element tends to zero in the limit,
then transferred momentum $q$ goes to zero.
To be convinced, let us write the quantity $N_h$ as
\ba
M_h=\frac{1}{s\beta}(q-q_\bot)^{\nu} J_{\nu}(q)=
-\frac{1}{s_0}q_\bot^\nu \bar{u}(p') O_\nu(q) u(p).
\ea
So this fact is the sequence of gauge invariance of current.

Here and further we restrict ourselves only to the approximation of a
"large logarithm" (i.e. the so-called Weizs\"acker-Williams approximation).
To do this, we perform the
integration over the transferred momentum using the relation
\ba
\int\frac {d^2\vec{q}\,\,\vec{q}_i\vec{q}_j}{\pi(\vec{q}^2+m_e^2\beta^2)^2}=\frac{1}{2}\delta_{ij}(L-1), \nn
\ea
where $L$ is the "large logarithm":
\ba
L=\ln\frac{Q^2s^2}{m_e^2s_0^2}\approx\ln\frac{s^2}{m_e^2M^2}\approx 23.
\ea

Writing the matrix element of the pion photo-production
sub-process $\gamma(q)+p\to N\pi_1\pi_2$ as
\ba
M_h^i=\frac{1}{s_0}\bar{u}(p')\hat{O}_i u(p),
\qquad
i=\pi^+\pi^-, \pi^+\pi^0,
\ea
and using the density matrix of a polarized proton as
$u(p,a)\bar{u}(p,a)=(\hat{p}+M)(1-\gamma_5\hat{a})$, where 4-vector
$a$ is the polarization vector of the initial proton (i.e. $p\cdot a = 0$),
we obtain for the differential cross section:
\ba
\frac{d\sigma_i}{dx_1dx_2} =\sigma_0\frac{S_i}{s_0^2}\frac{dx_1dx_2 d^2\vec{q}_1d^2\vec{q}_2}{xx_1x_2\pi^2}, \nn \\
\sigma_0=\frac{\alpha^2G^2}{64M^2\pi^3}(L-1),
\label{CrossSection}
\ea
where
\ba
S_i=\frac{M^2}{4}Tr(\hat{p}'+M)\vec{O}(\hat{p}+M)(1-\gamma_5\hat{a})
\vec{O}^*.
\ea
For HERMES conditions the factor $\sigma_0$ has a
rather large value $\sigma_0\approx 2\cdot 10^{-30}\cm^2$, i.e. of an order of two micro-barns.
For ZEUS detector the expected cross section is about three micro-barn.

\section{Matrix element of $\pi^+\pi^-$ production on a proton}
\label{SecPM}

The matrix element of the $\pi_+\pi_-$ meson production can be presented as a sum of
two separately gauge-invariant contributions
\ba
\bar{u}(p')O^{\pi^+\pi^-}u(p)=M_{\pi_+\pi_-}=M_1+M_2.
\label{MatrixElementPM1}
\ea
The first term describes the subprocess of creation of a pion pair by (virtual) photon and
(virtual) $\rho$-meson
\ba
M_1&=&\frac{1}{q_1^2-M_\rho^2}\brs{\bar{u}(p')\gamma^\rho u(p)} \times\nn\\
&\times&
\brf{
    \frac{\br{\br{2 q_--q},e}\br{-2q_++q_1}_\rho}{(q-q_-)^2-m^2}+
    \frac{\br{2q_--q_1}_\rho \br{\br{-2 q_++q},e}}{(q-q_+)^2-m^2}
    -2e_\rho
},
\ea
where $e=e(q)$ is the polarization vector of virtual photon,
$q_1=q_++q_--q$.
The second term describes
the subprocess of Compton scattering on a proton with emission of $\rho$-meson
which subsequently decays into a pion pair
\ba
M_2=\frac{1}{q_2^2-M_\rho^2+i\Gamma_\rho M_\rho}
\brs{
    \bar{u}(p')
    \brf{
        \hat{e}\frac{\hat{p}'-\hat{q}+M}{(p'-q)^2-M^2}\hat{v}+
        \hat{v}\frac{\hat{p}+\hat{q}+M}{(p+q)^2-M^2}\hat{e}
    }u(p)
},
\ea
where $q_2=q_++q_-,v=-q_++q_-$.

Both contributions ($M_1$and $M_2$)
satisfy the gauge condition: they turn to zero when we use the
mass shell conditions and replace $e(q)\to q$.

The matrix element (\ref{MatrixElementPM1}) can be written in the form
\ba
M_{\pi_+\pi_-}=\frac{s}{s_0}\brs{\bar{u}(p')\hat{O}_{+-}u(p)},
\label{AmplitudePM1}
\ea
where we left only quadratic over $\vec{q}$ terms and operator
$\hat O_{+-}$ has the form:
\ba
\hat{O}_{+-}= A\hat{q}_++B\hat{q}_-+C\hat{q}_\bot+D\hat{v}+
E\hat{q}_\bot\hat{p}_1\hat{v}+F\hat{v}\hat{p}_1\hat{q}_\bot,
\ea
where
\ba
A=\frac{4x^2x_+}{s_0D_\rho}\frac{\vec{q}\vec{q}_-}{M^2}; \qquad
B=\frac{4x^2x_-}{s_0D_\rho}\frac{\vec{q}\vec{q}_+}{M^2}; \qquad
D=\frac{2x_-^2x_+^2}{s_0D_\rho}\frac{\vec{p}'\vec{q}_-}{M^2}; \nn
\ea
\ba
C=\frac{2x}{D_\rho}; \qquad
E=\frac{x_+x_-}{sxD_\pi}; \qquad
F=\frac{x_+x_-}{sD_\pi}, \nn
\ea
and the $\rho$-meson and pion propagator denominators are
\ba
D_\rho&=&x\mu^2+\nu^2(1-x)^2+\chi_1+\chi_2+2\chi, \nn \\
D_\pi&=&-\mu^2(1-i\gamma_\rho)x_+x_-+x_1^2\chi_2+x_2^2\chi_1-2x_+x_-\chi+\nu^2(1-x)^2, \nn
\ea
\ba
\gamma_\rho=\frac{\Gamma_\rho}{M_\rho}, \qquad
\mu=\frac{M_\rho}{M}, \qquad \nu=\frac{m}{M}. \nn
\ea

\section{Matrix element of $\pi^+\pi^0$ production on a proton}
\label{SecP0}

For the subprocess $\gamma^*(q)+p(p)\to n(p')\pi_+(q_+)\pi_0(q_0)$
the amplitude $M_{+0}$ contains besides the Compton-like amplitude
\ba
M_1=\frac{1}{(q_++q_0)^2-M_\rho^2+i\Gamma_\rho M_\rho}
\brs{\bar{u}(p')(\hat{q}_0-\hat{q}_+)\frac{\hat{p}+\hat{q}+M}{(p+q)^2-M^2}\hat{e}u(p)},
\ea
also three terms corresponding to the conversion of the virtual photon and the virtual
$\rho$-meson to a pion pair:
\ba
M_2 &=&\frac{\bar{u}(p')\gamma_\rho u(p)}{\br{(q_++q_0)^2-M_\rho^2}\br{(q_++q_0-q)^2-M_\rho^2}}
\nn \\
&\times&\brs{e v_+\br{-q_+-q_0-q}_\rho+
v_{+\rho}\br{2q_++2q_0-q}e+e_\rho v_+\br{2q-q_+-q_0}}, \nn \\
M_3&=&-\frac{\bar{u}(p')(-\hat{q}_++\hat{q}_0+\hat{q}) u(p) e(-2q_++q)}
{\br{(q_++q_0)^2-M_\rho^2}\br{(q_+-q)^2-m^2}};\nn \\
M_4&=&\frac{\bar{u}(p')\hat{e} u(p)}{(q_++q_0-q)^2-M_\rho^2},
\qquad
v_+ = -q_+ + q_0.
\nn
\ea
It can be checked that the sum
\ba
M_{+0} = M_1 + M_2 + M_3 + M_4,
\ea
obeys the gauge condition. Writing this amplitude
in the form similar to (\ref{AmplitudePM1}) leads us to
\ba
M_{+0}=\frac{s}{s_0}\brs{\bar{u}(p')\hat{O}_{+0}u(p)},
\ea
where:
\ba
\hat{O}_{+0}&=&-\frac{\hat{v}_+\hat{p}_1\hat{q}_\bot}{sD_\pi}+
\frac{1}{D_{\rho} D_{\pi}}
\brs{
    -2\beta\hat{p}_1 \vec{q}\vec{v}_{+} +
    2\hat{v}_+\vec{q}(\vec{q}_{+} +\vec{q}_{0})+s_0(x_+ -x_0)\hat{q}_{\bot}
}
-\nn\\
&-&\frac{1}{s_0 D_{\rho} x_+}
\brs{s_0 x_+\hat{q}_\bot+4\vec{q}\vec{q}_+\hat{q}_0},
\ea
where the expressions for $s_0$, $D_\rho$, $D_\pi$ can be obtained from the
relevant expressions for process of $\pi_+\pi_-$ production (see Section~\ref{SecPM})
by replacement $\vec{q}_-,x_-\to \vec{q}_0,x_0$.

\section{Results}
\label{SectionResults}

The differential cross section of the pion pair electroproduction processes
we are considering here can be obtained using Eq.~(\ref{CrossSection}).
In the case of $\pi^+\pi^-$ pair production we have to insert the quantity
$S_{+-}$ into this formula (\ref{CrossSection}). This quantity has the form:
\ba
S_{+-}&=&(1-P_\pm)\left\{\frac{-2\frac{1}{M}\brs{\vec{q_+}\vec{a}}_z \cdot Im D_{\pi}}{D_{\rho}|D_{\pi}|^2 s_0} A+
\frac{4 Re D_{\pi}M^2}{s_0^2 D_{\rho}|D_{\pi}|^2} D \right\} + \nn \\
&+&(1+P_\pm)\left\{\frac{2M^2}{D_{\rho}^2 s_0^2}C
+\frac{2M^2}{s_0^2 |D_{\pi}|^2} F\right\},
\label{Spm}
\ea
where the permutation operator $P_\pm$ acts on any function
$F\br{x_+,x_-,\vec{q}_+,\vec{q}_-}$ in the following manner
\ba
P_\pm F\br{x_+,x_-,\vec{q}_+,\vec{q}_-}=F\br{x_-,x_+,\vec{q}_-,\vec{q}_+}
\ea
and the quantities $A$, $C$, $D$, $F$ in (\ref{Spm}) are given in Appendix~\ref{AppendixA}.

In the case of $\pi^+\pi^0$ pair production we have to insert the quantity
$S_{+0}$ into this formula (\ref{CrossSection}). This quantity has the form:
\ba
S_{+0}&=&-\frac{2Im D_\pi}{s_0 D_\rho^2|D_\pi|^2}
\brs{\frac{\brs{\vec{q}_+,\vec{a}}_z}{M}
(A_1+A_2D_\rho)+\frac{\brs{\vec{q}_0,\vec{a}}_z}{M}B}+\frac{2}{s_0 D_\rho^2}C+ \nn \\
&+&\frac{2}{D_\rho|D_\pi|^2}D+\frac{4Re D_\pi}{s_0 D_\rho^2|D_\pi|^2}E+\frac{2}{D_\rho|D_\pi|^2}F+
\frac{4Re D_\pi}{D_\rho|D_\pi|^2}G+\frac{2}{|D_\pi|^2}H,
\label{Sp0}
\ea
where quantities $A_1$, $A_2$, $B$, $C$, $D$, $E$, $F$, $G$, $H$
are given in Appendix~\ref{AppendixB}.

The Dalitz-distribution (the distribution over final pion energy fractions)
can be written in the following form:
\ba
\frac{d\sigma}{dx_1 dx_2} =
\sigma_0 F_i^{\text{unp}}\br{x_1,x_2},
\ea
where the function $F_i^{\text{unp}}\br{x_1,x_2}$
\ba
F_i^{\text{unp}}\br{x_1,x_2}
=
\int\frac{d^2\vec{q}_1}{\pi}\frac{d^2\vec{q}_2}{\pi}xx_1x_2\frac{S_i}{s_0^2}.
\label{FUnpDefForTable}
\ea
In the case of $\pi^+\pi^-$ production this function
is presented in Table~\ref{TablePM}; while in the
case of $\pi^+\pi^0$ production, in Table~\ref{TableP0}.
\begin{table}
  \begin{tabular}{|c|c|c|c|c|c|c|}
  \hline
    $x_+$ / $x_-$ & $0.2$   & $0.3$   & $0.4$   & $0.5$   & $0.6$   & $0.7$  \\
  \hline
    $0.2$ &         $0.268$ & $0.359$ & $0.362$ & $0.312$ & $0.250$ & $0.138$  \\
  \hline
    $0.3$ &         $0.337$ & $0.613$ & $0.708$ & $0.620$ & $0.410$ & \\
  \hline
    $0.4$ &         $0.351$ & $0.712$ & $0.899$ & $0.808$ &         & \\
  \hline
    $0.5$ &         $0.315$ & $0.625$ & $0.792$ &         &         & \\
  \hline
    $0.6$ &         $0.240$ & $0.406$ &         &         &         & \\
  \hline
    $0.7$ &         $0.137$ &         &         &         &         & \\
  \hline
  \end{tabular}
  \caption{The function $F_{+-}^{\text{unp}}\br{x_+,x_-}$
  (defined in (\ref{FUnpDefForTable})) is presented for different
  values of the final pion energy fractions $x_+$ and $x_-$
  for the $\pi^+\pi^-$ production case.}
  \label{TablePM}
\end{table}
\begin{table}
  \begin{tabular}{|c|c|c|c|c|c|c|}
  \hline
    $x_+$ / $x_0$ & $0.2$    & $0.3$    & $0.4$   & $0.5$   & $0.6$   & $0.7$  \\
  \hline
    $0.2$ &         $56.325$ & $23.089$ & $2.897$ & $0.418$ & $0.052$ & $0.003$  \\
  \hline
    $0.3$ &         $23.166$ & $14.824$ & $5.237$ & $0.421$ & $0.016$ & \\
  \hline
    $0.4$ &          $2.876$ &  $5.063$ & $0.990$ & $0.071$ &         & \\
  \hline
    $0.5$ &          $0.415$ &  $0.419$ & $0.067$ &         &         & \\
  \hline
    $0.6$ &          $0.051$ &  $0.016$ &         &         &         & \\
  \hline
    $0.7$ &          $0.003$ &          &         &         &         & \\
  \hline
  \end{tabular}
  \caption{The function $F_{+0}^{\text{unp}}\br{x_+,x_0}$
  (defined in (\ref{FUnpDefForTable})) is presented for different
  values of the final pion energy fractions $x_+$ and $x_0$
  for the $\pi^+\pi^0$ production case.}
  \label{TableP0}
\end{table}

\begin{figure}
\includegraphics[width=0.8\textwidth]{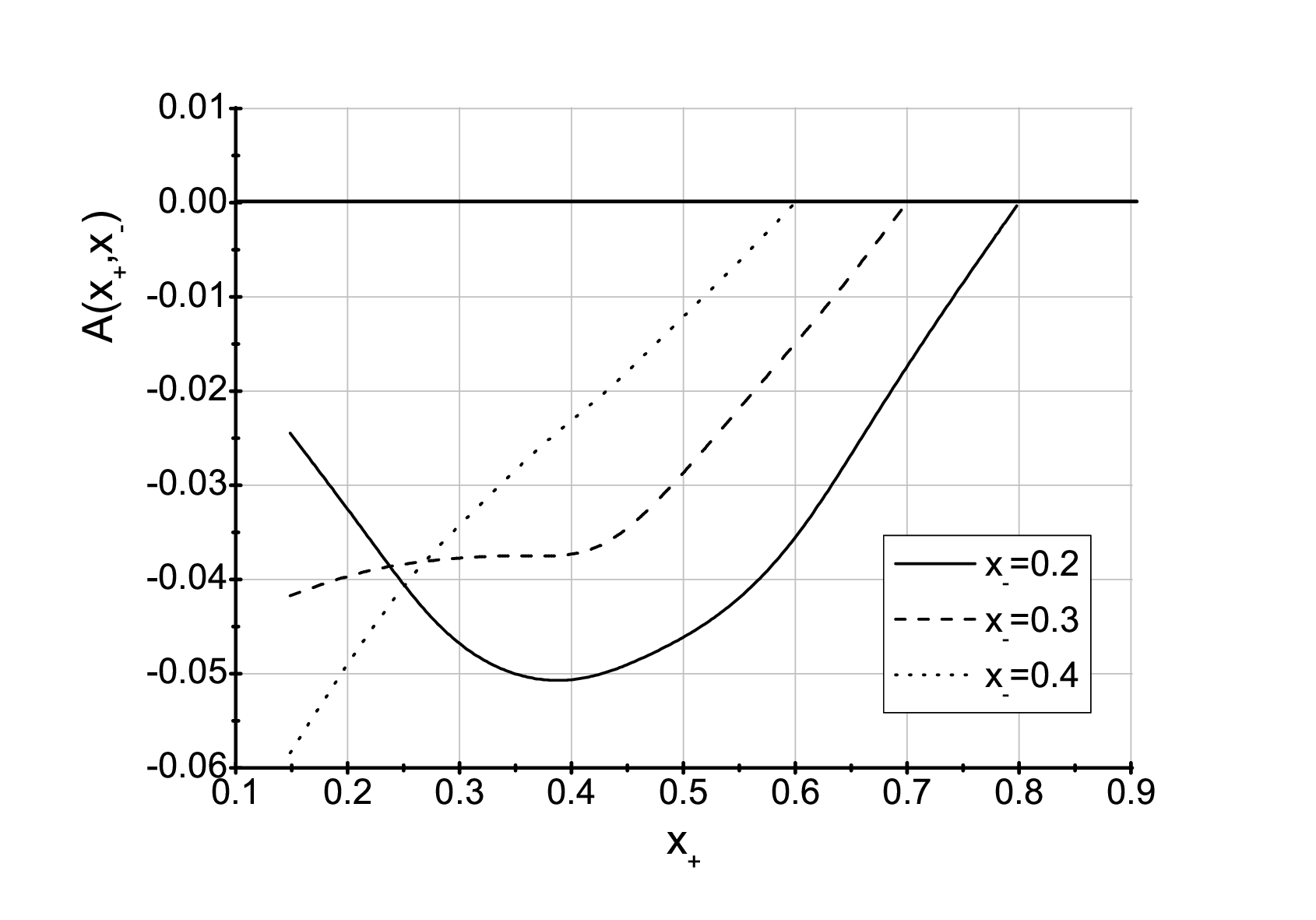}
\caption{
The asymmetry $A\br{x_+,x_-}$ (defined in (\ref{AsymmetryDef}))
as a function of $x_+$ with $x_-$ fixed to definite values.
}
\label{FigPM}
\end{figure}

\begin{figure}
\includegraphics[width=0.8\textwidth]{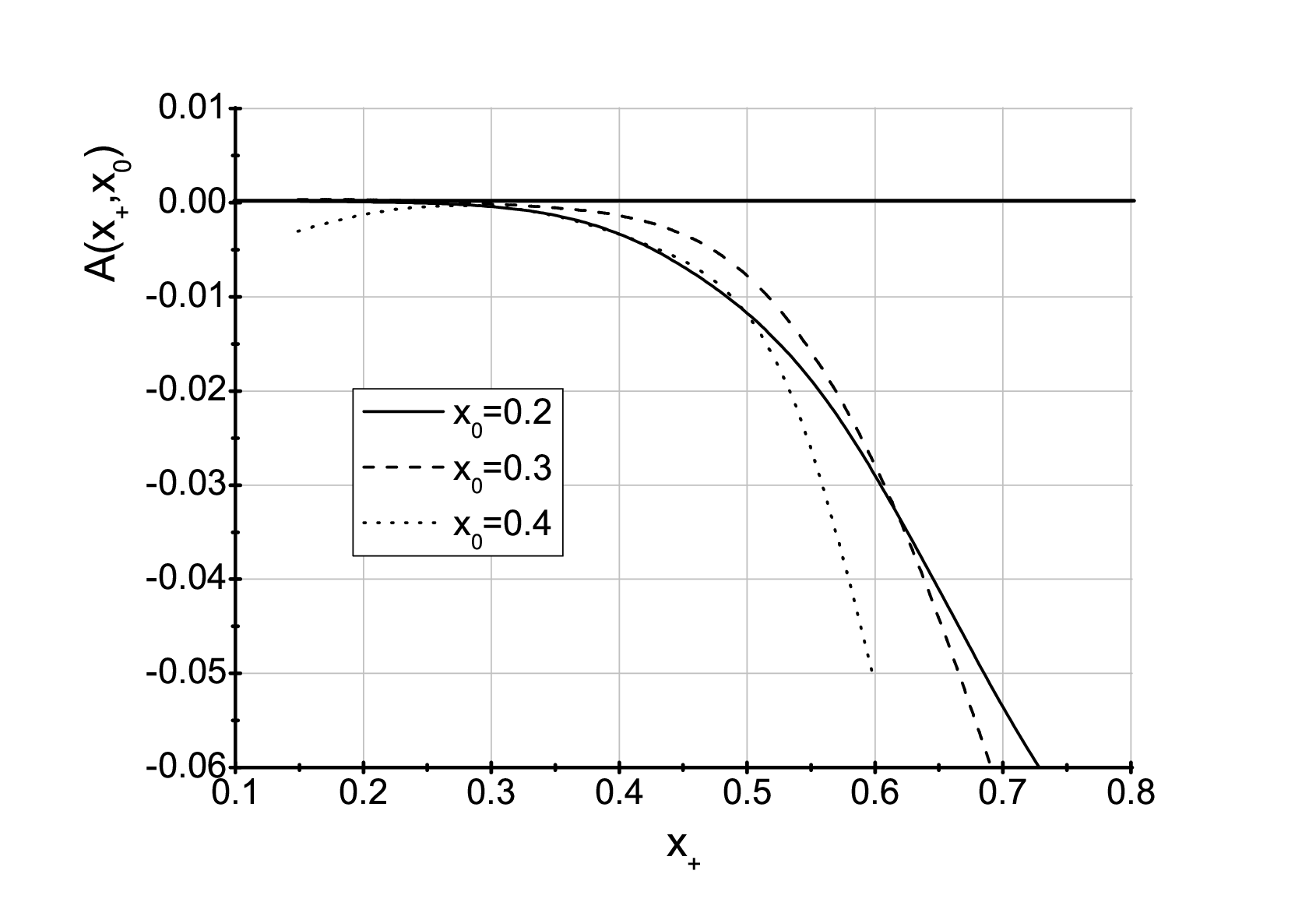}
\caption{
The asymmetry $A\br{x_+,x_0}$ (defined in (\ref{AsymmetryDef}))
as a function of $x_+$ with $x_0$ fixed to definite values.
}
\label{FigP0}
\end{figure}

In the experimental setup when we fix the azimuthal angle $\psi$
between the proton polarization vector $\vec{a}$ and the
transverse momentum of one of the pions $\vec{q}_+$,
i.e. $\psi=\br{\vec{a},\vec{q_+}}$, some polarization dependent
contributions appear:
\ba
    \int d\brm{\vec{q_1}} \int \frac{d^2 q_2}{\pi}
    x x_1 x_2 \frac{S_i}{s_0^2}
    =
    \brm{\vec a} \sin\psi ~ F_i^{\text{pol}}\br{x_1,x_2}
    +
    F_i^{\text{unp}}\br{x_1,x_2}.
\ea
And thus in experiment the asymmetry \cite{Sivers1,Sivers2}
\ba
    A\br{x_1,x_2}
    =
    \frac{F_i^{\text{pol}}\br{x_1,x_2}}
    {F_i^{\text{unp}}\br{x_1,x_2}},
    \label{AsymmetryDef}
\ea
can be measured. This asymmetries for both channels are presented
in Figs.~\ref{FigPM}, \ref{FigP0} as a function $x_+$ for fixed
values of $x_-$ or $x_0$.

We should note that the meaning of the quantities
$x$, $x_1$, $x_2$ is different in the laboratory reference frame and in
the center of mass of the initial particles (c.m. frame).
In the c.m. frame they are the energy fractions
of the jet consisting of the recoil proton and pions which obey the conservation laws
\ba
x+x_1+x_2=1, \qquad x_i=\frac{2E_i}{\sqrt{s}}, \qquad \frac{2m}{\sqrt{s}} < x_i < 1,
\qquad \frac{2M}{\sqrt{s}}<x<1.
\ea

In the laboratory frame, keeping in mind the explicit form of the light-like vectors
$\tilde{p}$ and $\tilde{p}_1\approx p_1$:
\ba
\tilde{p}=\frac{M}{2}(1,-1,0,0), \qquad p_1=E(1,1,0,0), \nn
\ea
we have for energies:
\ba
E_i=\frac{M}{2}x_i+\frac{\vec{q}_i^2+m^2}{2Mx_i},\qquad
E'=\frac{M}{2}x+\frac{(\vec{q}_1+\vec{q}_2)^2+M^2}{2Mx},\nn
\ea
\ba
M+\frac{s_1}{2Mxx_1x_2}=E_1+E_2+E', \qquad \frac{m}{M}<x_i<1. \nn
\ea
The values of 3-momenta squares are
\ba
{\bf q}_i^2&=&\vec{q}_i^2+\brs{-\frac{M}{2}x_i+\frac{\vec{q}_i^2+m^2}{2Mx_i}}^2, \nn \\
{\bf p}^2&=&\br{\vec{q}_1+\vec{q}_2}^2+\br{-\frac{M}{2}x+\frac{(\vec{q}_1+\vec{q}_2)^2+M^2}{2Mx}}^2.
\nn
\ea
We specifically emphasize that the left part of these equalities has the
3-dimensional vectors and the right part has the two-dimensional vectors
(their components perpendicular to the initial electron direction are zero).

The Dalitz-plot distributions for the unpolarized case (see Fig.~\ref{FigPM} and \ref{FigP0},
while $\sigma_0 \approx 2 \mu b$) seem to be accessible in experiments of HERMES.
The one-spin asymmetries (see (\ref{AsymmetryDef})) have a value of several per cent and
can be measured as well.
A preliminary comparison of these quantities is in qualitative agreement with data obtained at HERMES
\cite{Nagaytsev}.

\section{Conclusion}
\label{Conclusion}
In this study, we examined the production of pions in the electron-proton scattering,
taking into account that the proton in the initial state is polarized.
In the calculation, we used the HERMES energy.

In this paper, we investigated the differential spectral distribution of the
energy fractions $x_{\pm,0}$ and spin asymmetry in the region $x_+ + x_- < 1$,
and $x_+ + x_0 < 1$. Our results can be used in HERMES and COMPASS experiments.

\appendix

\section{The production of $\pi^+\pi^-$ pair}
\label{AppendixA}
Here we present the explicit expressions for the values of $A$, $C$, $D$, $F$ for
the $\pi_+\pi_-$ production (see (\ref{Spm})):
\ba
A=A_1 \chi_+ +\frac{1}{2}A_3 \chi_- +\frac{1}{2}A_4 s_1 +\frac{1}{2}A_5;
\ea
\ba
C=C_1 \chi_{+}^2 \chi_ {-}+\frac{1}{2}C_3 \chi_+\chi_-\chi +C_4 \chi_+^2 +\frac{1}{2}C_6 \chi^2 +
\frac{1}{2}(C_7 +s_1 C_8) \chi_+ \chi_- + \nn \\
(C_9 +s_1 C_{10})\chi_+ \chi + (C_{13}s_1 +C_{14} +C_{15}
s_1^2)\chi_+ +\frac{1}{2}(C_{19}+C_{20}s_1^2)\chi +
\frac{1}{2}C_{21}s_1^2 +\frac{1}{2}C_{22}s_1^3; \ea
\ba
D=D_1\chi_{+}^2\chi_- +D_3 \chi_+ \chi_{-}^2 +D_5 \chi_{+}^2 \chi
+\frac{1}{2}D_7 \chi_{+}\chi_{-}\chi +
D_8\chi_{+}^2 +\frac{1}{2}\chi^2 (D_{10}+s_1 D_{11}) + \nn \\
\frac{1}{2}\chi_+\chi_- D_{12}+
(D_{13}+s_1 D_{14})\chi_+\chi +(D_{17}s_1 +D_{18} +s_1^2 D_{19})\chi_+ + \nn \\
\frac{1}{2}(D_{23}s_1 +D_{24}+D_{25}s_1^2)\chi
+\frac{1}{2}D_{26}s_1^2 +\frac{1}{2}D_{27}s_1^3;
\ea
\ba
F=F_1 \chi_+^3 +F_3\chi_+^2 \chi_- +F_5 \chi_+ \chi^2 +F_7
\chi_+^2 \chi +\frac{1}{2}F_9 \chi_+ \chi_- \chi +
\chi_+^2(F_{10}s_1 +F_{11}) + \nn \\
\frac{1}{2}\chi^2(F_{14}s_1 +F_{15}) +\frac{1}{2}\chi_+ \chi_- (F_{16}s_1 +F_{17}) +
(F_{18}s_1 +F_{19})\chi_+\chi + \nn \\
(F_{22}s_1 +F_{23}+s_1^2 F_{24})\chi_+ +
\frac{1}{2}(F_{28}+s_1 F_{29})\chi +\frac{1}{2}s_1^2 F_{30} +\frac{1}{2}s_1^3 F_{31}.
\ea
The expressions for the coefficients are
\ba
A_1=-x(1-x)(1-x_-)x_-^3; \,\,A_2=xx_+x_-(1-x)(2x_+-x_-(1-x_+)); \nn \\
A_3=2x(2-x_-)x_+x_-^2(1-x); \,\,A_4=x_-^2(1-x)(1+x_+-x_-); \nn \\
A_5=xx_-^3x_+(1-x)(x_+-x_-). \ea
\ba
C_1&=&-\frac{2}{x_+}x^4(x_-^2-x_+(1-x_-)); C_3=4x^4(2-x); C_4=-\frac{2}{x_+}x^4x_-^4; \nn \\
C_6&=&4x^3; C_7=-2x^4[x_+x_-(1-x)-2(x_+^2+x_-^2)]; C_8=\frac{2x^3}{x_+x_-}(1-x)^2; \nn \\
C_9&=&4x_-^2x^4(1+x_-); C_{10}=-\frac{2}{x_+}x_-x^3; C_{13}=\frac{2}{x_+}x_-^2x^3(1-x); \nn \\
C_{14}&=&2x_-^4x^4; C_{15}=-\frac{x^2}{x_+}; C_{19}=4x_-^2x_+^2x^4; \nn \\
C_{20}&=&\frac{2}{x_+x_-}x^2(1-x); C_{21}=-x^2(1-x);
C_{22}=\frac{x}{x_+x_-}.
\ea
\ba
D_1=2(xx_-)^2 x_+;
D_3=2(xx_-x_+)^2; D_5=x^2x_-^2x_+(1-x_-); \nn \\
D_7=x_+x_-x^2(x_--x_+)(2-x);  D_8=x^2x_-^4x_+(1+x_-); \nn \\
D_{10}=-2(xx_+x_-)^2(1-x)(x_--x_+); D_{11}=xx_+x_-(1+x)(x_--x_+);
\nn \\ D_{12}=-3(xx_-x_+)^2(x_--x_+); \nn \\
D_{13}=x_-^2x^2x_+[x_-^3+x_-^2(1-2x_+) -x_-x_+(1+x_+)+2x_+^2]; \nn
\\
D_{14}=\frac{1}{2}xx_-[x_-^3-x_-^2+x_-x_+(6-x_+)+x_+(3x_+-2)];
D_{17}=-xx_-^3x_+;  \nn \\D_{18}=-x^2x_+^2x_-^4(x_--x_+);
D_{19}=\frac{1}{2}x_-[x_+-xx_-]; \nn \\
D_{23}=\frac{1}{2}xx_-x_+(1-x)(1-3x)(x_--x_+);
D_{24}=-(xx_-x_+)^2(x_-^2+x_+^2)(x_--x_+); \nn \\
D_{25}=\frac{1}{2}x(1-x)(x_--x_+);
D_{26}=-\frac{1}{2}x_-x_+(1-x)(x_--x_+); \nn \\
D_{27}=\frac{1}{2}(x_--x_+).
\ea
\ba
F_1=\frac{1}{2}x_-^4x_+^2(1-x_-); F_3=-\frac{1}{2}x_-^3x_+^2[x_-^2-x_-(1-x_+)+x_+(2+x_+)];  \nn \\
F_5=2x_-^4x_+^3;
F_7=x_+^2x_-^4(1+x_+-x_-); F_9=-2(x_-x_+)^3; \nn \\
F_{10}=-\frac{1}{2}x_-^3x_+^2(2-x_-);
F_{11}=-\frac{1}{2}x_-^4x_+^3[x_-(3-x)-2x_+]; F_{14}=-2(x_-x_+)^3; \nn \\
F_{15}=2(x_-x_+)^4(1-x);
F_{16}=\frac{1}{2}(x_-x_+)^2[x_-^2+x_+^2+2(1-x)]; \nn \\
F_{17}=-\frac{1}{2}(x_-x_+)^3[(1-x)(x_-^2+x_+^2)+2(x_- -x_+)^2];
F_{18}=x_-^3x_+^2(x_--x_+); \nn \\
F_{19}=-x_-^4x_+^3(3-x)(x_--x_+); F_{22}=(x_-x_=)^3(x_--x_+);
F_{23}=\frac{1}{2}(x_-x_+)^4(x_--x_+); \nn \\
F_{24}=\frac{1}{4x}x_-^2x_+[-x_-^2(1-x_-)-x_+^2(3-x_-)+2x_+(1+x_-^2)];
F_{28}=(x_-x_+)^4(x_--x_+)^2; \nn \\
F_{29}=-\frac{1}{2x}(x_-x_+)^2[1+x^2];
F_{30}=-\frac{1}{4x}(x_-x_+)^2(1-x)(x_--x_+)^2; \nn \\
F_{31}=\frac{1}{4x} x_-x_+(x_--x_+)^2.
\ea

\section{The production of $\pi^+\pi^0$ pair}
\label{AppendixB}

Here we present the explicit expressions for the scalar coefficients in trace for the
$\pi_+\pi_0$ production (see (\ref{Sp0})):
\ba
A_1=s_1[\chi_+a_1+\chi_0 a_2+\chi a_3]+s_1^2a_4; \nn \\
A_2=\chi_+ a_5+\chi_0 a_6+\chi a_7+s_1 a_8+a_9,
\ea
with
\ba
a_1=-\frac{1}{x_+}x^2x_0^3; a_2=x_+x_0x^2; a_3=4(x_0x)^2; a_4=\frac{1}{x_+}xx_0^2; \nn \\
a_5=x^2x_0^4; a_6=x_0x_+x^2(x_+-x_0+x_+x_0); a_7=-2x_+x_0^3x^2, \nn \\
a_8=x(x_+-x_0); a_9=x_+x_0^3x^2(x_+-x_0).
\ea
\ba
B=s_1(b_1\chi_++b_2\chi_0+b_3\chi)+b_4s_1^2,
\ea
with
\ba
b_1=-(x_0x)^2; b_2=(xx_+)^2;b_3=-4x_+x_0x^2; b_4=-xx_+.
\ea
\ba
C=c_1\chi_+^2\chi_0+c_2\chi_0^2\chi_++c_3\chi_+\chi_0\chi+
c_4\chi_+^2+(c_5+s_1c_6)\chi_+\chi_0+(c_7+s_1c_8)\chi_+\chi+ \nn \\
(c_9s_1^2+s_1c_{10}+c_{11})\chi_+c_{12}s_1^2\chi_0+(s_1c_{13}+c_{14})s_1\chi+ \nn \\
+s_1^2c_{15}+s_1^3c_{16},
\ea
with
\ba
c_1=-\frac{2x_0^2(x_+ +x_0 -1)^4}{x_+},\,c_2=-2(x_+ -1)(x_+ +x_0 -1)^2, \,c_3=4x_0(x_++x_0-1)^4, \nn \\
c_4=-\frac{2x_0^4(x_++x_)-1)^4}{x_+},\,c_5=-2(x_+-2)x_0^2(x_++x_0-1)^4, \nn \\
c_6=-\frac{(x_++x_0-1)^3(x_++2x_0)}{x_+},\,c_7=4x_0^3(x_++x_0-1)^4,\,
c_8=\frac{x_0(x_+ +x_0 -1)^3}{x_+}, \nn \\
c_9=-\frac{(x_+ +x_0 -1)^2}{4x_+},\,c_{10}=-\frac{x_0^2(x_++x_0 -1)^3(x_++2x_0)}{x_+},\,
c_{11}=2x_0^4(x_++x_0-1)^4, \nn \\
c_{12}=-\frac{(x_++x_0-1)^2}{4x_0},\,c_{13}=\frac{(x_++x_0-1)}{x_+},\,c_{14}=x_+x_0(x_++x_0-1)^3, \nn \\
c_{15}=-\frac{1}{4}(x_++x_0-1)^2(x_++x_0),\,c_{16}=-\frac{x_++x_0-1}{4x_+x_0}.
\ea

\ba
D=d_1\chi_+^3+d_2\chi_0^3+d_3\chi_+^2\chi_0+d_4\chi_0^2\chi_++d_5\chi_+^2\chi+ \nn \\
d_6\chi_0^2\chi+d_7\chi^2\chi_++d_8\chi^2\chi_0+d_9\chi_+\chi_0\chi+ \nn \\
(s_1d_{10}+d_{11})\chi_+^2+(s_1d_{12}+d_{13})\chi_0^2+(s_1d_{14}+d_{15})\chi^2+ \nn \\
(s_1d_{16}+d_{17})\chi_+\chi_0+(s_1d_{18}+d_{19})\chi_+\chi+(s_1d_{20}+d_{21})\chi_0\chi+ \nn \\
(s_1^2d_{22}+s_1d_{23}+d_{24})\chi_++(s_1^2d_{25}+s_1d_{26}+d_{27})\chi_0+ \nn \\
(s_1^2d_{28}+d_{29})\chi_+s_1^2 d_{30}+s_1^3 d_{31},
\ea
with
\ba
d_1=-\frac{1}{2}x^2 x_0^2(x_0-1),\,d_2=-\frac{1}{2}x^2 x_+^2(x_+-1), \nn \\
d_3=-\frac{1}{2}x^2 x_0(x_0^2+(x_+-1)x_0+x_+(x_++2)), \nn \\
d_4=-\frac{1}{2}x^2 x_+(x_0^2+(x_+ +2)x_0+x_+(x_+-1)), \nn \\
d_5=x^2 x_0^2 (x_+-x_0+1)x_0^2,\,d_6=-x^2 x_+^2(x_+-x_0-1), \nn \\
d_7=2x^2 x_+x_0^2, \,d_8=2x^2 x_+^2x_0,\,d_9=-2x^2 x_+x_0,
d_{10}=\frac{1}{2}x x_0(x_++2x_0 -2), \nn \\
d_{11}=-\frac{1}{2}x^2 x_+x_0^2(x_+(x_0-2)+x_0(x_0+2)),
d_{12}=\frac{1}{2}x x_+(2x_+ +x_0-2),\nn \\
d_{13}=-\frac{1}{2}x^2 x_+^2x_0(x_+(x_++2)+x_0(x_0-2)),
d_{14}=-2x x_+x_0, \,d_{15}=2x^2 x_+^2x_0^2(1-x), \nn \\
d_{16}=-x((x_+-1)x_0 -x_+), \nn \\
d_{17}=-\frac{1}{2}x^2 x_+x_0 (x_0^3+(x_++2)x_0^2+(x_+ -4)x_+x_0 +x_+^2(x_+ +2)), \nn \\
d_{18}=-x x_0(x_+ -x_0), \,d_{19}=x^2 x_0^2 x_+(x_+ -x_0)(3-x), \nn \\
d_{20}=x x_+(x_+ -x_0),\,d_{21}=-x^2 x_+^2 x_0(x_+ -x_0)(3-x), \nn \\
d_{22}=\frac{1}{4}\left(-\frac{x_0^2}{x_+}-3x_+ +2 \right), \,
d_{23}=-\frac{1}{2}x x_+x_0(x_+ -x_0)(2+x), \nn \\
d_{24}=\frac{1}{2}x^2 x_+^2 x_0^2(x_+-x_0)^2, \,
d_{25}=\frac{1}{4}\left(-\frac{x_+^2}{x_0}-3x_0 +2 \right), \nn \\
d_{26}=-\frac{1}{2}x_+ x x_0(x_+ -x_0)(x -1), \,
d_{27}=\frac{1}{2}x^2 x_+^2 x_0^2(x_+ -x_0)^2, \nn \\
d_{28}=-x, \,d_{29}=x^2x_+^2(x_+ -x_0)^2, \,d_{30}=-\frac{1}{4}(x_+ -x_0)(1-x), \nn \\
d_{31}=\frac{(x_+-x_0)^2}{4xx_+x_0}.
\ea

\ba
E=e_1\chi_+^2\chi_0+e_2\chi_0^2\chi_++e_3\chi_0^2\chi+e_4\chi^2\chi_0+e_5\chi_+\chi_0\chi+
e_6\chi_+^2 \nn \\
(s_1e_7+e_8)\chi^2+(s_1e_9+e_{10})\chi_0\chi_++(s_1e_{11}+e_{12})\chi\chi_++ \nn \\
(s_1e_{13}+e_{14})\chi\chi_0+(s_1^2e_{15}+s_1e_{16}+e_{17})\chi_++(s_1e_{18}+e_{19})s_1\chi_0+ \nn \\
(s_1^2e_{20}+s_1e_{21}+e_{22})\chi+s_1^2 e_{23}+s_1^3 e_{24},
\ea
with
\ba
e_1=-x_0 x^3(x_0+1), \,e_2=-x_+ x^3(x_+-1), \,e_3=-x_+x^3(x_+-1), \nn \\
e_4=-2x^3 x_+ x_0, \,e_5=-x x_0(2x_+ -x_0-1), \,e_6=x^3 x_0^3(x_0+1), \nn \\
e_7=2x^2 x_0, \, e_8=-2x^3x_+x_0^3, \,e_9=-x^2, \,e_{10}=xx_+x_0(x_++(x_+-2)x_0), \nn \\
e_{11}=\frac{x_0 x^2(2x_+-x_0)}{2x_+},\,e_{12}=-x^3x_0^3(2x_+ -x_0-1),\,
e_{13}=-\frac{1}{2}x^2(3x_+ +2x_0 -2), \nn \\
e_{14}=x_+x_0x^3(x_++(x_+-2)x_0),\,e_{15}=\frac{x(x_+-x_0)}{4x_+},\,
e_{16}=-\frac{1}{2}x_0^2x^2(x_0+2), \nn \\
e_{17}=x_+x_0^3x^3(x_+-x_0),\,e_{18}=\frac{x(x_+-x_0)}{4x_0},\,e_{19}=\frac{1}{2}x^2x_+^2x_0, \nn \\
e_{20}=-\frac{x(x_+ -x_0)}{2x_+},\,e_{21}=x^2x_0(x_+^2-x_0^2+x_0),\,e_{22}=x^3x_+x_0^3(x_+-x_0), \nn \\
e_{23}=\frac{1}{4}x(x_+ -x_0)(1-x),\,e_{24}=\frac{1}{4}\left(\frac{1}{x_+}-\frac{1}{x_0}\right).
\ea

\ba
F=f_1\chi_+^2+f_2\chi_0^2+f_3\chi^2+f_4\chi\chi_0+f_5\chi_+\chi_0+
f_6\chi_0\chi + \nn \\
(s_1f_7+f_8)\chi_++(s_1f_9+f_{10})\chi_0+s_1\chi f_{11}+s_1^2f_{12},
\ea
with
\ba
f_1=-\frac{1}{2}x_+x_0^2x(x_0-1),\,f_2=-\frac{1}{2}x_+^2x_0 x(x_+-1),\,f_3=2x_+^2x_0^2x, \nn \\
f_4=-\frac{1}{2}x_+x_0 x(x_+^2 +x_0^2 +1-x),\,f_5=x_+xx_0^2(x_+-x_0),\,f_6=-x_+^2x_0x(x_+-x_0), \nn \\
f_7=\frac{1}{2}x_0(x_0^2+(x_+ -1)x_+),\,f_8=\frac{1}{2}x_+^2x_0^2 x(x_+ -x_0),\,
f_9=\frac{1}{2}x_+(x_+^2+(x_0 -1)x_0), \nn \\
f_{10}=-\frac{1}{2}x_+^2x_0^2 x(x_+-x_0), \,f_{11}=x_+x_0x,\,f_{12}=\frac{1}{4}x_0^2x.
\ea

\ba
G=g_1\chi^2+g_2\chi_0\chi_++g_3\chi\chi_++g_4\chi\chi_0+ \nn \\
(s_1g_5+g_6)\chi_++s_1\chi_0g_7+(s_1g_8+g_9)\chi+s_1^2g_{10},
\ea
with
\ba
g_1=-x_+x_0^2x^2,\,g_2=\frac{1}{2}x_+x_0x^2, \,g_3=\frac{1}{2}x_0^3x^2,\,
g_4=\frac{1}{2}(x_+-1)x_+x_0x^2,\,g_5=\frac{1}{4}x_0^2x, \nn \\
g_6=\frac{1}{4}x_+x_0^3x^2,\,g_7=-\frac{1}{4}x_+^2x,\,g_8=\frac{1}{2}(x_+-x_0)x_0x,\,
g_9=-\frac{1}{2}x_+x_0^3x^2,\,g_{10}=\frac{1}{4}(x_+-x_0).
\ea

\ba
H=h_1\chi_++h_2\chi_0+h_3\chi+h_4s_1+h_5,
\ea
with
\ba
h_1=-\frac{1}{4}x_+x_0^2(x_0^2 +x_+(x_0-1)),\,h_2=-\frac{1}{4}x_+^2 x_0(x_+^2+(x_+-1)x_0),\,
h_3=-\frac{1}{2}x_+^2x_0^2x, \nn \\
h_4=\frac{x_+(x_+ -x_0)^2 x_0}{4x},\,h_5=\frac{1}{4}x_+^2x_0^2(x_+ -x_0)^2.
\ea


\end{document}